\documentclass[conference]{IEEEtran}
\IEEEoverridecommandlockouts
\usepackage{cite}
\usepackage{amsmath,amssymb,amsfonts}
\usepackage{algorithmic}
\usepackage{graphicx}
\usepackage{textcomp}
\usepackage{xcolor}
\usepackage{subfig}
\usepackage{listings}
\usepackage{balance}
\def\BibTeX{{\rm B\kern-.05em{\sc i\kern-.025em b}\kern-.08em
    T\kern-.1667em\lower.7ex\hbox{E}\kern-.125emX}}

\usepackage[most]{tcolorbox}

\usepackage{enumerate}
\usepackage{setspace}

\usepackage{xpatch}
\usepackage{todonotes}
\usepackage[acronym, toc]{glossaries}
\definecolor{darkred}{rgb}{0.5,0,0}
\definecolor{darkgreen}{rgb}{0,0.5,0}
\definecolor{darkblue}{rgb}{0,0,0.5}
\definecolor{gray}{rgb}{0.35,0.35,0.35}

\usepackage[pagebackref=false,
            colorlinks=true,
            linkcolor=gray,
            bookmarks=true,
            filecolor=gray,
            urlcolor=gray,
            citecolor=gray
           ]{hyperref}
\usepackage[nameinlink]{cleveref}
\usepackage{paralist}
\usepackage{xurl}
\usepackage{multirow}
\usepackage{tabularx}
\usepackage{fontawesome}

\newacronym{IoT}{IoT}{Internet of Things}
\newacronym{IIoT}{IIoT}{Industrial Internet of Things}
\newacronym{IETF}{IETF}{Internet Engineering Task Force}
\newacronym{MUD}{MUD}{Manufacturer Usage Description}
\newacronym{ENISA}{ENISA}{European Network and Information Security Agency}
\newacronym{FTD}{FTD}{Full Thread Device}
\newacronym{MTD}{MTD}{Minimal Thread Device}
\newacronym{REED}{REED}{Router Eligible End Device}
\newacronym{SED}{SED}{Sleepy End Device}
\newacronym{MED}{MED}{Minimal End Device}
\newacronym{FED}{FED}{Full End Device}
\newacronym{MLE}{MLE}{Mesh Link Establishment}
\newacronym{DoS}{DoS}{Denial of Service}
\newacronym{DDoS}{DDoS}{Distributed Denial of Service}
\newacronym{SDN}{SDN}{Software Defined Networking}
\newacronym{IDS}{IDS}{Intrusion Detection System}
\newacronym{ACL}{ACL}{Access Control List}
\newacronym{JSON}{JSON}{JavaScript Object Notation}
\newacronym{ACE}{ACE}{Access Control Entry}
\newacronym{CLI}{CLI}{Command Line Interface}
\newacronym{RTOS}{RTOS}{Real Time Operating System}
\newacronym{RLOC16}{RLOC16}{Routing Locator}
\newacronym[longplural=Systems on Chip]{SOC}{SOC}{System on Chip}
\newacronym{RCP}{RCP}{Remote Co-Processor}
\newacronym{OMR}{OMR}{Off-Mesh Routable}
\newacronym{UDP}{UDP}{User Datagram Protocol}
\newacronym{SLAAC}{SLAAC}{Stateless Address Auto Configuration}
\newacronym{DHCPv6}{DHCPv6}{Dynamic Host Configuration Protocol version 6}
\newacronym{DHCP}{DHCP}{Dynamic Host Configuration Protocol}
\newacronym{TCP}{TCP}{Transmission Control Protocol}
\newacronym{ICMP}{ICMP}{Internet Control Message Protocol}
\newacronym{YANG}{YANG}{Yet Another Next Generation}
\newacronym{ULA}{ULA}{Unique Local Address}
\newacronym{ML-EID}{ML-EID}{Mesh-Local Endpoint Identifier}
\newacronym{EID}{EID}{Endpoint Identifier}
\newacronym{RA}{RA}{Router Advertisement}
\newacronym{NDP}{NDP}{Neighbor Discovery Protocol}
\newacronym{RTT}{RTT}{Round Trip Time}
\newacronym{RTT2}{RTT}{Real Time Transfer}
\newacronym{LLDP}{LLDP}{Link Layer Discovery Protocol}
\newacronym{YAML}{YAML}{Yet Another Markup Language}
\newacronym{UART}{UART}{Universal Asynchronous Receiver Transmitter}
\newacronym{URL}{URL}{Uniform Resource Locator}
\newacronym{API}{API}{Application Programming Interface}
\newacronym{HTTPS}{HTTPS}{HyperText Transfer Protocol Secure}
\newacronym{IP}{IP}{Internet Protocol}
\newacronym{IPv6}{IPv6}{Internet Protocol Version 6}
\newacronym{DNS}{DNS}{Domain Name System}
\newacronym{OVS}{OVS}{OpenVSwitch}
\newacronym{TLS}{TLS}{Transport Layer Security}   
\newacronym{6LoWPAN}{6LoWPAN}{IPv6 over Low-Power and Lossy  Networks}
\newacronym{CPS}{CPS}{Cyber-Physical Systems}
\newacronym{CI}{CI}{Critical Infrastructure}
\newacronym{SCADA}{SCADA}{Supervisory Control and Data Acquisition}
\newacronym{RPL}{RPL}{Routing Protocol for Low Power and Lossy Networks}
\newcommand{\name}{MeshGuard}

\definecolor{eclipseStrings}{RGB}{42,0.0,255}
\definecolor{eclipseKeywords}{RGB}{127,0,85}
\colorlet{numb}{magenta!60!black}

\lstdefinelanguage{json}{
    basicstyle=\normalfont\ttfamily,
    commentstyle=\color{eclipseStrings}, 
    stringstyle=\color{eclipseKeywords}, 
    numbers=left,
    numberstyle=\scriptsize,
    stepnumber=1,
    numbersep=2pt,       
    xleftmargin=1.5em, 
    showstringspaces=false,
    breaklines=true,
    frame=lines,
    string=[s]{"}{"},
    comment=[l]{:\ "},
    morecomment=[l]{:"},
    literate=
        *{0}{{{\color{numb}0}}}{1}
         {1}{{{\color{numb}1}}}{1}
         {2}{{{\color{numb}2}}}{1}
         {3}{{{\color{numb}3}}}{1}
         {4}{{{\color{numb}4}}}{1}
         {5}{{{\color{numb}5}}}{1}
         {6}{{{\color{numb}6}}}{1}
         {7}{{{\color{numb}7}}}{1}
         {8}{{{\color{numb}8}}}{1}
         {9}{{{\color{numb}9}}}{1}
}
\newcommand*\emptycirc[1][1ex]{\tikz\draw (0,0) circle (#1);} 

\newcommand*\fullcirc[1][1ex]{\tikz\fill (0,0) circle (#1);}

\begin{document}

\title{\name: MUD-Based Network Access Control for Large-Scale Thread-Powered IoT Networks
}

\author{
\IEEEauthorblockN{
Dominik Roy George\IEEEauthorrefmark{2}\IEEEauthorrefmark{1},
Wouter Van Hoof\IEEEauthorrefmark{2},
Habib Mostafaei\IEEEauthorrefmark{2},
and Savio Sciancalepore\IEEEauthorrefmark{2}
}
\IEEEauthorblockA{\IEEEauthorrefmark{2}\textit{Eindhoven University of Technology (TU/e)}, The Netherlands\\
d.r.george@tue.nl, w.p.h.j.v.hoof@student.tue.nl, h.mostafaei@tue.nl, s.sciancalepore@tue.nl}
\IEEEauthorblockA{\IEEEauthorrefmark{1}\textit{COSIC, KU Leuven}, Belgium\\
dgeorge@esat.kuleuven.be}
}

\maketitle

\begin{abstract}

\footnote{This is a personal copy of the authors. Not for redistribution. The final version of the paper will be available soon through the digital library IEEExplore.} The IETF standard Manufacturer Usage Description (MUD) enables manufacturers to equip IoT devices with certified URLs that provide traffic profiles for those devices, helping administrators enforce network access control. 
However, MUD assumes devices operate on full IP stacks and therefore does not account for constrained IoT devices running Thread--the dominant low-power mesh networking standard--which lacks complete TCP/IP functionality. While prior work proposes extensions to support MUD in Thread environments, these approaches are limited to simple topologies with a single border router and do not scale to realistic deployments with multiple, heterogeneous border routers.
We introduce MeshGuard, a framework enabling MUD-based access control in complex Thread networks, with any number of border routers. MeshGuard extends the Mesh Link Establishment (MLE) protocol to deliver MUD information from constrained devices to border routers regardless of network topology. Moreover, MeshGuard leverages Software-Defined Networking (SDN) to synchronize access control lists across all routers. Experiments on our proof-of-concept with real devices (nRF5340, nRF52833, Raspberry-Pi 3) demonstrate enhanced security, minimal overhead, and linear scalability compared to state-of-the-art approaches.

\end{abstract}


\hypersetup{
  linkcolor=gray,
  urlcolor=gray,
  citecolor=gray,
}




\glsresetall
\section{Introduction}
\label{sec:introduction}

With heterogeneous \gls{IoT} devices and systems increasingly pervading any aspect of our daily life, from home to transportation and industrial plants, concerns about their security and vulnerability to various cyberattacks are today greater than ever~\cite{iot-2022-2033},~\cite{fu2022_dsn},~\cite{Tagliaro2024_raid}. To name a few examples, in March 2025, the Shadowserver Foundation reported that more than 86,000 IoT devices were compromised by a new botnet named Eleven11~\cite{eleven11_2025}. Similarly, in November 2024, security researchers unveiled a widespread \gls{DDoS} campaign leveraging accessible tools and targeting IoT devices and enterprise servers through weak credentials and misconfigured services~\cite{matrix_2024}. 
These examples demonstrate the need for security mechanisms capable of identifying and rejecting traffic unnecessary to the basic functions of \gls{IoT} devices, which could lead to potential threats.

In this context, the \gls{IETF} recently published the \gls{MUD} standard specification~\cite{MUD-RFC}. The standard requires manufacturers of \gls{IoT} devices to equip their products with \emph{information} about their expected traffic profile, i.e., incoming and outgoing traffic that such devices exchange when integrated into user deployments. System administrators can use such traffic profiles together with additional local rules to deploy effective network access control, allowing the device to work as intended while preventing anomalous and unnecessary traffic~\cite{access2021_ramos},~\cite{morgese2022_acsac}.
With some manufacturers actively pushing the deployment of MUD profiles (e.g., Cisco~\cite{cisco-mud}), several works in the literature investigate the integration of such specifications into various application scenarios (see Sec.~\ref{sec:related-work} for an overview). 

However, the MUD standard defines mechanisms that allow only devices running the TCP/IP protocol stack to deliver MUD-related information. The standard provides no mechanism and neither guidelines for integration on more constrained IoT devices, unable to run over TCP/IP. These include Thread, i.e., the de-facto commercial standard protocol stack for such devices \textcolor{black}{in Smart Homes and Buildings}, used today in millions of commercial low-power IoT devices~\cite{ThreadSpec},~\cite{kim2019_commag}.
Recently, Houben et al.~\cite{thesis_luke_paper} addressed this gap by proposing a mechanism for Thread devices to share the location of the device's MUD traffic profile (i.e., the MUD \gls{URL}) with the Thread border router. 
While pioneering, this solution exhibits several limitations. \textcolor{black}{The approach is tailored to a simple Thread topology comprising one single \gls{FTD} directly connected to a single border router. }
In large-scale deployments with multi-hop mesh links and several border routers, several issues arise. First, the proposed extension of the \gls{MLE} protocol is incompatible with Minimal Thread Devices (MTDs). Second, the solution fails to reliably deliver the MUD URL to the entity enforcing network access control (namely, the MUD Manager) when multi-hop paths separate Thread end devices from the border router. Third, in networks with multiple border routers, the design lacks mechanisms to synchronize and maintain consistency of access control rules across border routers, both at first delivery and during updates.
Our results in this paper show that, although deploying such a state-of-the-art solution, adversaries can still exploit outdated and vulnerable protocols to compromise IoT devices, or can leverage physically captured nodes to mount attacks on Internet-connected hosts. Thus, current deployments still lack a comprehensive security architecture capable of enforcing effective MUD-based access control across large-scale Thread meshes with heterogeneous border routers.
These observations motivate our central research question:
\begin{tcolorbox}[colback=gray!10, colframe=gray!60!black, boxrule=0.5pt, arc=2pt, left=4pt, right=4pt, top=2pt, bottom=2pt]
To what extent can we enforce MUD-based network access control in heterogeneous large-scale Thread networks characterized by several border routers?
\end{tcolorbox}

\textcolor{black}{
{\bf Research Challenges.} Answering this research question requires solving several research challenges: (i) identifying the layer and protocol of the Thread stack most suitable for enabling automated delivery of MUD profiles from Thread-enabled IoT devices to the MUD manager, independently of the network topology (ii) designing a Thread-compliant solution that enables any Thread device, independently of its deployment, to reliably deliver such information, and (iii) designing and deploying a Thread-compliant network architecture that orchestrates the enforcement of MUD rules for a Thread-powered network independently from the specific Thread border router managing incoming and outgoing traffic. 
}

\textcolor{black}{
{\bf Contribution.}
In this paper, we propose \name, a lightweight, scalable, and dependable framework allowing to apply \gls{MUD}-based network access control in a constrained, heterogeneous and large-scale Thread-powered IoT network. 
Our work provides manyfold contributions:
\begin{itemize}
    \item We provide \name, a framework that extends MUD in two ways. First, \name\ defines a standard-compliant extension of the \gls{MLE} protocol, allowing any Thread node to reliably distribute \gls{MUD}-related information (URL) to the entity in charge of network access control (the MUD manager), independently of network size and topology. Moreover, \name\ applies the \gls{SDN} paradigm in the system design, to ensure consistency and synchronized management of access control rules among several constrained border routers, as required by large-scale Thread networks.
    \item  We implement \name\ in a real-world proof-of-concept using the Thread devices nRF5340 and nRF52833 and several Raspberry Pi embedded boards serving as border routers. 
    \item We report the results of several experimental tests evaluating the security and network overhead of \name. Our solution preserves the correct routing of allowed traffic while rejecting malicious incoming and outgoing connection attempts. It also adds negligible overhead to the latency experienced by constrained Thread devices, and scales linearly with the number of border routers.
\end{itemize}
}
This work makes a major step ahead towards enabling the deployment of MUD-based network access control in large-scale Thread-powered networks, significantly enhancing their security. \textcolor{black}{At the same time, \name\ guarantees effective application of MUD-based network security even in challenging low-power IoT networks, characterized by multiple Border Routers and several unstable hops separating end devices from them. Thus, \name\ ensures that the system performs reliably and securely independently from operational conditions, enhancing the overall dependability of the system. }

{\bf Roadmap.} This paper is organized as follows. 
Sec.~\ref{sec:related-work} reviews related work,
Sec.~\ref{sec:background} introduces the preliminaries, Sec.~\ref{sec:scenario} describes the considered scenario and adversarial model, Sec.~\ref{sec:design} provides the details of \name, Sec.~\ref{sec:implementation} describes our proof-of-concept, Sec.~\ref{sec:results} extensively evaluate \name\ experimentally on our proof-of-concept, Sec.~\ref{sec:discussion} discusses relevant aspects of \name, 
and finally, Sec.~\ref{sec:conclusion} concludes the paper and outlines future work.
\section{Related Work}
\label{sec:related-work}

\textcolor{black}{Several solutions in recent years have been proposed to prevent attacks to constrained \gls{IoT} devices by enforcing behavioral rules on network gateways.  
Early solutions consider networks based on the IEEE 802.15.4 communication technology which use the protocol \gls{RPL} at the network layer~\cite{verma2020_sensors}, and use behavioral rules mostly for intrusion detection~\cite{raoof2018_comst}. Some proposals focus on the defining network rules for detecting specific attacks, e.g., the sinkhole~\cite{sarawi2023_access} and the neighbor attacks~\cite{farzaneh2019_icwr}, while others aim to detect generic intrusions~\cite{raza2013_adhocj}.
None of these solutions is currently a standard, which limits their overall relevance to practitioners. Moreover, these solutions do not require the delivery of information (e.g., MUD URLs) to a management entity (e.g., the MUD Manager). Thus, they cannot help in integrating network access control based on MUD in Thread-powered networks.
At the same time, several other solutions consider more powerful TCP/IP-based IoT networks, e.g., based on IEEE 802.11 and WiFi~\cite{datta2025_cose},~\cite{islam2024_ucc}. The devices in such networks are less constrained, and use protocols different than the ones we focus in our work.
}

\textcolor{black}{After the standardization of \gls{MUD},} several works have explored integrating it into different \gls{IoT} networking environments, but they target architectures that differ from the Thread-based setting considered in this paper. We summarize their key characteristics in~\Cref{tab:related-work-comp} and briefly discuss them below.
\begin{table}[ht]
    \centering
    \caption{\textcolor{black}{Qualitative comparison of state-of-the-art \gls{MUD}-based schemes and our work. The symbol \fullcirc\ indicates that a particular feature is supported, while the symbol \emptycirc\ indicates that the feature is not supported.}
    }
    \label{tab:related-work-comp}
    \scalebox{0.75}{
    \color{black}
    \begin{tabular}{l|ccccc}
         Ref. & \begin{tabular}[c]{@{}c@{}}Constrained \\ Devices \end{tabular}  & \begin{tabular}[c]{@{}c@{}}Thread \\ Networks\end{tabular} & \begin{tabular}[c]{@{}c@{}}Multiple \\ Border \\ Routers \end{tabular}& \begin{tabular}[c]{@{}c@{}}MUD \\ orchestration\\ via SDN\end{tabular} &\begin{tabular}[c]{@{}c@{}}Tested \\ against\\ Real-World \\ Attacks\end{tabular}  \\
        \hline
        \cite{harish2022iot}     &  \emptycirc & \emptycirc & \emptycirc & \fullcirc & \emptycirc\\
        \cite{singh2019clearer} &  \fullcirc & \emptycirc & \emptycirc& \emptycirc & \emptycirc \\
        \cite{Hamza2018} &  \fullcirc & \emptycirc & \emptycirc & \fullcirc & \fullcirc \\
        \cite{Hamza2019} &   \fullcirc & \emptycirc & \emptycirc & \emptycirc &\fullcirc  \\
        \cite{hamza2022_iotj} & \fullcirc & \emptycirc & \emptycirc & \fullcirc & \fullcirc \\
        \cite{hamza2022_tdsc} & \fullcirc & \emptycirc & \emptycirc & \emptycirc & \emptycirc \\
        \cite{Krishnan2022} &  \fullcirc & \emptycirc & \emptycirc & \emptycirc & \fullcirc  \\
        \cite{Ranganathan2019_NIST} &  \fullcirc & \emptycirc &\emptycirc & \fullcirc & \fullcirc \\
        \cite{thesis_luke_paper} & \fullcirc  & \fullcirc & \emptycirc & \emptycirc &\fullcirc \\
        \hline
        {\bf Ours} & \fullcirc & \fullcirc & \fullcirc & \fullcirc & \fullcirc\\
        \hline
    \end{tabular}
    }
\end{table}
Since \gls{MUD} has been conceived for the \gls{IoT}, many scientific proposals take into account by design its integration on constrained devices featuring limited energy, storage, and computation~\cite{singh2019clearer},~\cite{Hamza2018},~\cite{Hamza2019},~\cite{hamza2022_iotj},~\cite{hamza2022_tdsc},~\cite{Krishnan2022},~\cite{Ranganathan2019_NIST}, and~\cite{thesis_luke_paper}. 
Although not taking into account such constraints explicitly, proposals like the ones of Harish et al.~\cite{harish2022iot}, Hamza et al.~\cite{Hamza2018},~\cite{hamza2022_iotj}, and Ranganathan et al.~\cite{Ranganathan2019_NIST} recognized the value of \gls{SDN} for MUD orchestration and proposed preliminary architectures integrating such technologies. Many of them also tested the security of the proposed architecture using real-world attacks, such as Mirai malware, \gls{DDoS}, and packet flooding, to name a few, typically by checking that regular traffic is allowed while malicious one is effectively blocked by the MUD manager.


\textcolor{black}{Existing approaches for MUD mostly focus on WiFi network environments; only the seminal work in~\cite{thesis_luke_paper} addresses highly constrained Thread deployments.} The framework in~\cite{thesis_luke_paper} enables Thread devices to convey their MUD URLs to a MUD manager, allowing the corresponding policies to be enforced at the edge, i.e., on the Thread border router. However, the design has two fundamental limitations. First, their MUD-URL delivery mechanism assumes that all Thread devices are within direct radio range of the Border Router. This assumption represents an unrealistic constraint for Thread meshes, that may contain hundreds of nodes arranged in complex multi-hop topologies~\cite{siliconLabs}. Second, the solution supports only a single Border Router. Therefore, it cannot operate in large-scale deployments where multiple heterogeneous Border Routers are the norm. Enforcing MUD policies in such environments requires dedicated coordination mechanisms across these routers. When evaluated under these more realistic conditions (see Sec.~\ref{sec:results}), the proposal in~\cite{thesis_luke_paper} fails to prevent attacks on the Thread network. These shortcomings motivate the design and analysis undertaken in this paper.

\section{Preliminaries}
\label{sec:background}

\subsection{Thread
}\label{sec:openthread}
Thread is a low-power, low-latency wireless mesh networking protocol stack integrating several open and well-known standard protocols for low power and lossy networks, including: (i) Constrained Application Protocol (CoAP) for the application layer, (ii) \gls{UDP} for the transport
layer, (iii) \gls{6LoWPAN} for the IP adaptation layer, (iv) \gls{MLE} for network joining and maintenance, and (v) IEEE 802.15.4 for the PHY and MAC layers. 
Among short-range wireless options (e.g., Bluetooth), Thread is often preferred in energy-constrained environments because it offers very low power consumption and extended device battery life~\cite{ieee802154, thread-low-power-value}.
We describe below some relevant features, useful for fully grasping the rationale of our work. 

\medskip
\noindent{\bf Node Types.} In line with IEEE 802.15.4, nodes in a Thread network can be either \glspl{FTD}, if they continuously have their radio turned on, or \glspl{MTD}, if they are usually in sleep mode to save energy and occasionally wake up and poll for existing messages~\cite{thread-networking-fundamentals, openthread-node-types}. While \glspl{MTD} can only work as end devices, \glspl{FTD} can take any \gls{REED} role, i.e., Thread Router and Border Router. 

\medskip
\noindent{\bf Border Routers and Thread Leader.} Thread networks can include several Thread Border Routers, providing a connection to external networks (including the Internet). When several border routers operate in the network, they autonomously elect a Thread Leader, i.e., a border router responsible for managing router identifiers and tracking network configurations, e.g., the available \gls{OMR} prefixes, the addresses of the border routers, and the external routes~\cite{openthread-node-types}. 


\medskip
\noindent{\bf Joining a Thread network.} To join a Thread network, new devices establish a connection to a device already part of the Thread network through \gls{MLE}~\cite{thread-networking-fundamentals, openthread-node-types}. According to the specification:  (i) the end device requests to join by sending a message \emph{MLE Parent Request}  to a default multicast IP address, used to discover the neighboring routers and \glspl{REED} within the transmission range in the Thread network; (ii) all \glspl{REED} respond with a message \emph{MLE Parent Response}, containing the leader data; (iii) the end device sends a message \emph{Child ID Request} to the selected preferred parent to establish a link to it; (iv) the parent node responds with a message \emph{Child ID Response} to confirm that a link has been established, with the response containing the new \gls{RLOC16} address of the child, and possibly the current network data.


\medskip
\noindent{\bf Thread Routing.} \gls{RLOC16} addresses of a device change with topology changes. Thus, they cannot be used to uniquely identify a node over time. To this aim, Thread devices maintain several other \gls{IPv6} unicast addresses, namely \glspl{EID} \cite{thread-networking-fundamentals}: link-local (on a single-hop level), mesh-local (within the Thread network), and global (outside the Thread network)~\cite{openthread-addressing}. 
Thread networks maintain a mesh-local prefix. Using this prefix, each device generates its \gls{ML-EID} using the protocol \gls{SLAAC} \cite{thread-networking-fundamentals, rfc7217} as a replacement for \gls{DHCP}. To translate an \gls{EID} to a routable \gls{RLOC16}, the \glspl{FTD} maintain a EID-to-RLOC mapping. 
Parent nodes also use \glspl{EID} to buffer messages for sleepy child nodes.


For routing packets to and from external networks, the border router registers an \gls{OMR} prefix and an external route to facilitate bidirectional communication. The Thread leader then updates the network data and spreads the update to the rest of the network. Once a device receives this update, it can locally compute an \gls{OMR} \gls{IP} address. The \gls{OMR} prefix must be a routable \gls{IPv6} prefix with which devices outside of the Thread network can reach the Thread devices. 


When a new link is established, the parent node provides \emph{network data} to the newly-joined devices, including an \gls{OMR} IP prefix. 
The \gls{IP} address can be obtained from the OMR prefix either via \gls{DHCPv6} or \gls{SLAAC}. \gls{DHCPv6} requires a DHCP server, while for \gls{SLAAC} the child generates a random unique address itself and, if it is a \gls{MTD}, shares it with its parent through \gls{MLE} Update Request and Response messages. They are also used for liveness checks, data synchronization, and reconnection. 

\medskip
\noindent{\bf OpenThread.} OpenThread is an open-source implementation of Thread, provided by Google to promote its adoption and improve interoperability~\cite{openthread, kim2019_commag}. As this is the only open source implementation of Thread, used also in Industry, we use the terms OpenThread and Thread interchangeably.

\glsreset{MUD}
\subsection{Manufacturer Usage Description}
\label{sec:mud}

The \gls{MUD} standard, defined in RFC 8520 and developed by the \gls{IETF}, provides a framework for a device to automatically specify the necessary network access it requires to function properly \cite{RFC8520}. 

According to MUD, manufacturers specify inbound and outbound network services used by a given device in a file, namely
\emph{\gls{MUD} File}, containing a \gls{JSON} serialized \gls{YANG} model for \glspl{ACL} \cite{rfc8519, rfc7951}. 
Typical MUD files include the \gls{MUD} version, the last date on which the file was modified, a signature file for validation, the \textit{from-device-policy} (outbound traffic), and the \textit{to-device-policy} (inbound traffic). 
The MUD file is uploaded by the manufacturer on a server, namely the \emph{\gls{MUD} Server}, reachable publicly at a given URL, namely the \emph{\gls{MUD} URL}. To enable the retrieval of the MUD file corresponding to a given device, the manufacturer deploys the MUD URL in the non-volatile memory of the IoT device, so that the device can provide it at boot time to the local network administrator who will deploy it.
\gls{MUD} provides three options for a device to share its \gls{MUD} URL: (i) through \gls{DHCP}, (ii) through \gls{LLDP}, or (iii) securely via an X.509 extension. None of these options is available in very constrained Thread-based networks.

\glsreset{SDN}
\subsection{Software Defined Networking and OpenFlow}
\label{sec:sdn}
{\bf\gls{SDN}.} Traditional IT networks operate with switches and routers provided by different vendors, each with its own configuration tools and software stacks, being hardly reconfigurable on the fly and integrable for network administrators~\cite{improving-net-managment}. The \gls{SDN} paradigm aims to address this problem by providing a three-layer network architecture that unifies device management procedures while abstracting hardware differences~\cite{improving-net-managment}. At the \emph{physical layer}, also known as the data plane, SDN switches simply forward messages based on tables of rules specifying a \emph{flow}, i.e., a set of packet values to match on and certain actions to take. 
The data plane exposes to the higher layer the \textit{southbound \gls{API}}, used to manage the data forwarding devices directly.
At the \emph{control layer} (or control plane) a specific entity, namely the SDN Controller, is tasked with configuring the devices in the data plane. Similarly to an operating system, it manages the low-level details of the underlying physical devices and presents a consistent interface to enable high-level configuration of such devices. The control plane exposes to the higher layer the \textit{northbound \gls{API}}, which allows configuration changes. 
At the highest layer, the \emph{Management Layer} (or management plane) consists of network applications that implement the high-level control logic of the network. Examples include load-balancers, network monitoring services, and firewalls. 

\medskip
\noindent{\bf OpenFlow.} 
OpenFlow is the \gls{SDN} protocol that defines both the data-plane switch architecture and the \textit{southbound \gls{API}} used for controller--switch communication \cite[Chapter~5]{goransson2016software}.
In OpenFlow, incoming packets are processed based on \textit{flows}. An OpenFlow switch contains a number of \textit{flow tables}, a \textit{group table}, a \textit{meter table}, \textit{channels} to controllers, and input and output \textit{ports}. A flow table holds a number of \textit{flow entries}. A flow entry has \textit{match fields}, \textit{counters}, and \textit{instructions} that must be performed on matching packets. The switch can receive packets through physical or logical ports. Upon reception of a packet, the contents of the packet are matched in order against each entry of the first flow table. Upon a match, the packet is processed according to the instruction, i.e., modifications, update, forward, drop, or output to a port.
The group table contains \textit{action buckets} with multiple actions to perform if a packet is directed to a group. Finally, the meter table stores any required statistics per flow. For implementing a firewall, the switch can be instructed to drop all matching packets according to some \gls{ACL}. 

The OpenFlow protocol specifies three types of communication between switches and controllers: controller-to-switch, asynchronous, and symmetrical. As the name suggests, controller-to-switch messages are sent by the controller to the switch and possibly require a reply. This includes messages to update the state of all tables inside the switch, as well as any other configuration settings. 
Asynchronous messages are sent by the switch to the controller without the controller initiating communication. This type of messages mainly includes status messages. It also includes the \textit{packet\_in} message, which is used to forward packets for which no flow rule is present to the controller. In turn, the controller most likely responds with an update to the flow tables to process such packets in the future or with a \textit{packet\_out} message, which instructs the switch to output the included packet onto a specific port. All communication between switches and controllers occurs over a dedicated channel, running on top of \gls{TCP} or \gls{TLS}. 


\section{Scenario and Adversary Model}\label{sec:scenario}

\noindent{\bf Scenario.} 
We consider a wireless \gls{IoT} network spanning a medium-range geographical area (e.g., a building) and including several constrained \gls{IoT} devices provided by various manufacturers for various tasks. These devices feature limited computational, storage, communication, and energy resources.
All \gls{IoT} devices, including the border routers and internal routers in the local network, run the protocol stack Thread discussed in Sec.~\ref{sec:background}.
Thus, wireless communication between devices occurs over the communication frequency 2.4~GHz using the IEEE~802.15.4 communication technology. 
\textcolor{black}{The \gls{IoT} devices in the local network are connected in a mesh network topology to several border routers, allowing to avoid single points of failure and facilitating the reach to outside networks, e.g., the Internet~\cite {kim2019_commag}.}
Every new device joins the existing Thread network through the \gls{MLE} protocol. 
We also consider that the network is run by a single network administrator, willing to enforce access control policies on incoming and outgoing traffic based on the \gls{MUD} specification. In this context, the presence of a constrained wireless technology and the deployment of several border routers make MUD deployment in this scenario particularly challenging. Our solution, namely \emph{\name}, allows us to tackle these challenges and deploy MUD-based access control policies on network traffic efficiently and effectively.

\medskip
\noindent{\bf Adversarial Model.} 
We consider an adversary willing to get access to the IoT devices and use them to attack Internet hosts.
On one hand, we consider a fully remote attacker willing to gain illegitimate access to the \gls{IoT} devices through brute-forcing of credentials associated with services available by default but practically unused on such devices. An example of such a threat is the behavior of the early malware Mirai~\cite{antonakakis2017understanding}, brute-forcing Telnet credentials on port 23.
On the other hand, our adversary model also envisions an attacker featuring physical access to the IoT devices, who gains a foothold on the device via physical tampering, wireless hacking via vulnerable Thread protocols, or manual connection. Once onboard the device, the attacker could use it as a weapon, e.g., in a \gls{DDoS} attack~\cite{dedonno}. 
In this context, our solution aims to thwart attempts to (i) compromise IoT devices from the Internet, and (ii) prevent the usage of physically-compromised devices to attack Internet hosts. 
To this end, our solution incorporates concepts from the \gls{SDN} paradigm, assuming that all SDN components--controllers and switches--are fully trusted and uncompromised. This trust is maintained through proper network segmentation, which prevents devices from establishing remote sessions with these critical elements. Finally, our adversary does not possess the manufacturer's private key and, therefore, cannot forge legitimate \gls{MUD} files.
\section{The Design of \name}
\label{sec:design}

\textcolor{black}{
\subsection{Design Requirements}
\label{sec:req}
We tackle the overarching research objective of applying behavior-based network access control based on the IETF standard MUD in multi-hop, multi-border router, low-power IoT networks, particularly the ones using the emergent Thread protocol stack. These networks do not run the TCP/IP protocol stack; thus, the solutions recommended by the MUD standard to allow IoT devices to deliver the URLs of their MUD profiles to the MUD manager are not applicable in this context.
}

Designing a solution to address such a challenge requires meeting several key requirements. \textit{First}, the system must enable any Thread device to convey its MUD URL to the network’s access-control authority (the MUD Manager) at join time, regardless of the device's role or the number of network hops separating it from a Thread Border Router.
Such a solution should also require minimal additional overhead, and integrate straightforwardly with existing network joining protocols. \textit{Second}, we need to consider that real-world large-scale Thread networks count several border routers. Such border routers should enforce MUD-based network access control consistently, using updated and synchronized network access control rules. Maintaining the synchronization of MUD-related information on the border router should also be lightweight, and scale reasonably with the size of the network. \textit{Third}, enforcing network access control rules must enhance security while imposing minimal overhead in terms of added latency. 
These requirements guide the design of \name, described in detail below.

\subsection{Network Architecture}
\label{sec:architecture}

Our framework, namely \name, enforces MUD-based network security in a constrained multi-hop mesh IoT network running on Thread and characterized by multiple Thread border routers. 
\Cref{fig:mudthread-v2-arch} presents the network architecture of \name, which includes: 
\begin{figure}[tp]
    \centering
    \includegraphics[width=0.9\columnwidth]{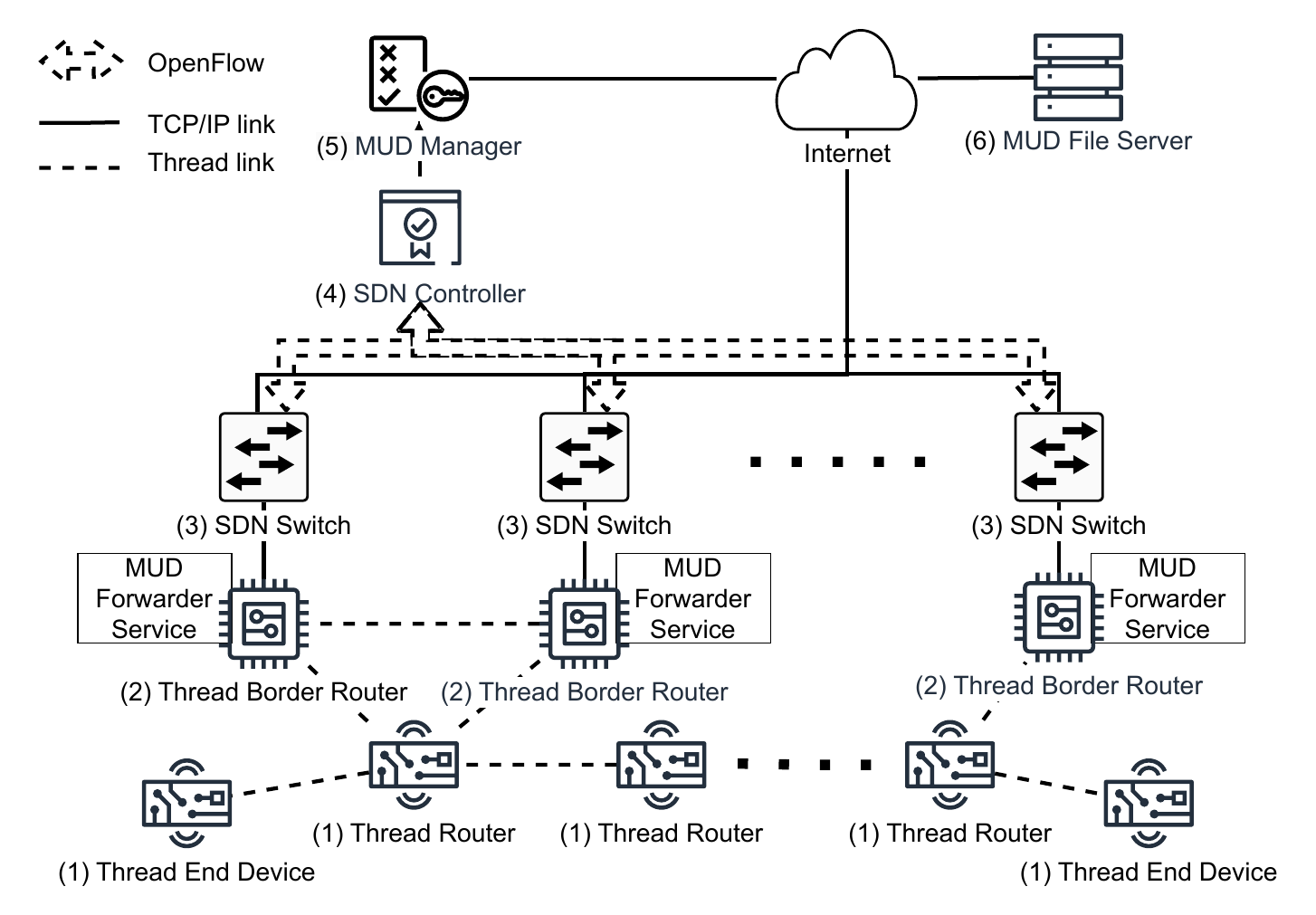}
    \caption{Architecture overview of \name.}
    \label{fig:mudthread-v2-arch}
\end{figure}
\begin{inparaenum}[(1)]
    \item the Thread network, consisting of several internal routers and end devices; 
    \item several Thread border routers, providing connection of the Thread network to the external network and hosting a \gls{MUD} Forwarder service, in charge of routing MUD-related traffic;
    \item several \gls{SDN} switches, co-located each with one Thread border router and routing traffic according to locally-stored flow tables;
    \item the \gls{SDN} Controller, that manages the forwarding tables stored within each \gls{SDN} switch and communicates with them via OpenFlow;
    \item the \gls{MUD} Manager, responsible for managing MUD-based access control policies for Thread devices; and finally,
    \item the \gls{MUD} Server, hosting the \gls{MUD} files for the Thread devices on the public Internet.
\end{inparaenum}

\name~defines the mechanisms that allows new nodes to provide their MUD URL to the network at joining time. 
New \gls{IoT} devices, configured as \gls{MTD} or \gls{FTD} based on their capabilities, join the Thread network using the \gls{MLE} protocol and, when connected to a parent node, deliver their MUD URL to the closest Thread border router. As explained later in Sec.~\ref{sec:joining-mud}, differently from the work in~\cite{thesis_luke_paper}, which uses one border router, \name\ works for multi-hop mesh networks by using a virtual multicast IP address grouping all Thread border routers. Moreover, since \gls{MLE} does not provide a default mechanism for \glspl{MTD} to deliver information to their parent node, \name\ extends standard \emph{MLE Update Responses} for delivering this information in a standard-compliant way. To allow sharing and synchronization of MUD information on all border routers in the Thread network, \name\ relies on the SDN paradigm. 
SDN switches are deployed alongside Thread Border Routers to steer traffic into and out of the Thread network, following flow rules distributed by a centralized SDN controller via OpenFlow.
In our architecture, the MUD Manager is deployed as an SDN application responsible for configuring and managing these flow rules. \textcolor{black}{Specifically, we leverage OpenVSwitch~\cite{openvswitch} to provide consistent and atomic synchronization of traffic rules, clear separation of control and data planes, and improved observability. Note that all these properties are difficult to achieve with direct manipulation of iptables via a lightweight control agent, which may result in transient inconsistencies.}
The following section reports the operations required to install new rules based on MUD files provided by IoT devices.

\subsection{\name\ Protocol Flow}
\label{sec:joining-mud}

\begin{figure*}[tp]
    \centering
    \includegraphics[width=0.95\textwidth]{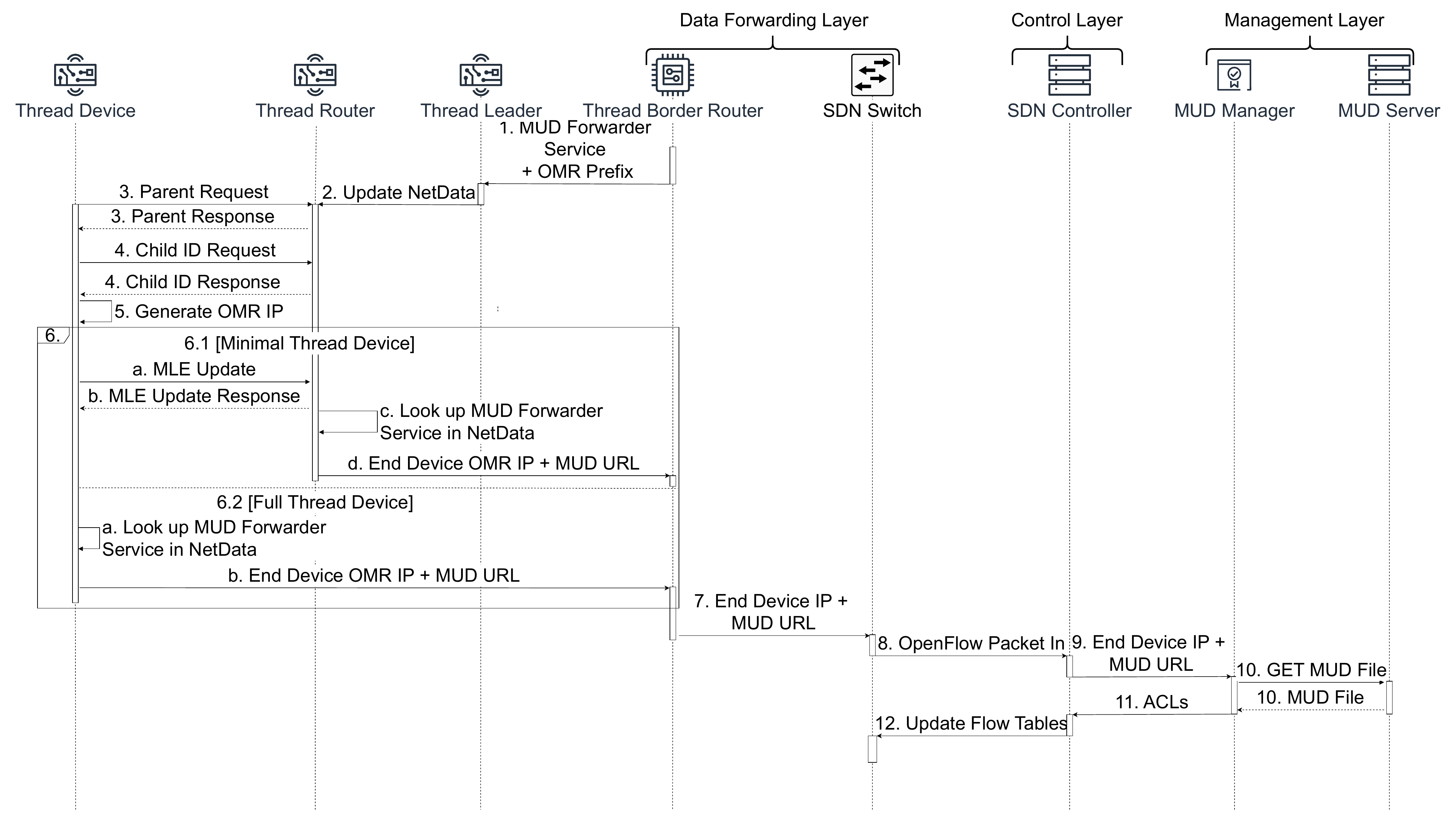}
    \caption{Sequence diagram of the operations required to allow a new IoT device to interact over the Internet while being protected via \name. }
    \label{fig:join-sequence}
\end{figure*}

\Cref{fig:join-sequence} provides the protocol flow allowing a newly-joined IoT device to interact properly over the Internet while being protected via \name. We discuss each step in more detail below. We refer the reader to background notions (Sec.~\ref{sec:background}) for the introduction of many technical notations used below.

\begin{enumerate}
    \item At network deployment time, each of the Thread border routers registers its \gls{MUD} Forwarder Service, an external route, and an \gls{OMR} prefix to the network data and shares such information with the Thread Leader.
    \item The Thread Leader distributes all such information as an update to the network data to all Thread Border Routers.
    \item When joining an existing Thread network, a Thread-enabled device (referred to as the \emph{joining device} from now on) multicasts a message \emph{\gls{MLE} Parent Request} message to all routers and \glspl{REED} in wireless range to obtain a parent node and join the network. Each of such devices replies, by standard, with a message \emph{\gls{MLE} Parent Response}.
    \item At the reception of MLE Parent responses, the joining device selects a preferred parent node based on local criteria and delivers to it a message \emph{Child ID request} to start the attach procedure. The preferred parent responds with a message \emph{Child ID Response}, containing the network data (see Sec.~\ref{sec:background}) useful to generate the IP address.
    \item The joining device extracts from the message the network data and generates its \gls{OMR} \gls{IP} address, which can be used to address it from external networks (e.g., Internet).

    \item The protocol flow evolves differently based on whether the joining device is a \gls{MTD} or a \gls{FTD}.
    \begin{enumerate}[]

        \item[6.1)] Case 1: the joining device is a \gls{MTD}:
        \begin{enumerate}
            \item Since \gls{MTD} devices cannot initiate communications, they have to rely on their respective parent nodes. Thus, the joining device delivers its \gls{MUD} \gls{URL} and \gls{OMR} \gls{IP} address to its preferred parent in a message, i.e., \emph{\gls{MLE} Update Request}.
            \item The parent node delivers a \emph{\gls{MLE} Update Response}. 
            \item The parent node looks up the available \gls{MUD} Forwarder Services in the received network data.
            \item  The parent node delivers to the \gls{IP} address of the chosen \gls{MUD} Forwarder Service an \gls{UDP} message containing the \gls{MUD} \gls{URL} and the \gls{IP} address of the (child) joining device.
        \end{enumerate}

        \item [6.2)] Case 2: the joining device is a \gls{FTD}:
        \begin{enumerate}
            \item The joining device looks up the available \gls{MUD} Forwarder Services in the received network data.
            \item  The joining device delivers to the \gls{IP} address of the chosen \gls{MUD} Forwarder Service an \gls{UDP} message containing its \gls{MUD} \gls{URL} and its \gls{IP} address.
        \end{enumerate}
        
    \end{enumerate}
    \item The UDP message is routed through the Thread network to a Thread Border Router that hosts an external route through which the destination \gls{IP} is reachable. This Thread border router forwards the message to its co-located \gls{SDN} switch.
    \item The \gls{SDN} switch extracts the \gls{MUD} information and encapsulates them into an OpenFlow \textit{packet\_in} message delivered to the \gls{SDN} Controller. 
    \item The \gls{SDN} controller delivers the \gls{IP} address of the joining node and the corresponding \gls{MUD} \gls{URL} to the \gls{MUD} Manager over the \textit{Northbound \gls{API}}. 
    \item If the \gls{MUD} file is not previously known or a newer version is available, the \gls{MUD} Manager fetches the \gls{MUD} file from the \gls{MUD} Server. 
    \item The \gls{MUD} Manager extracts the \glspl{ACL} from the \gls{MUD} file, and associates them to the \gls{IP} address of the joining device. Then, it sends the processed \glspl{ACL} to the \gls{SDN} controller over the \textit{Northbound \gls{API}}.
    \item The \gls{SDN} controller translates these new \glspl{ACL} into flow rules for the local flow tables, it updates such flow tables and distributes them to all \gls{SDN} switches via OpenFlow.
\end{enumerate}

At protocol completion, all \gls{SDN} switches enforce the same \gls{MUD}-based \glspl{ACL} for the newly-joined Thread device. As the traffic flowing through any Thread border router also passes through one of the \gls{SDN} switches, all gateways of the Thread network are protected by the same policies, ensuring the network-wide consistency of the policies used to secure network traffic.
\section{Proof of Concept} \label{sec:implementation}
\begin{figure}[tp]
    \centering
    \includegraphics[width=0.95\columnwidth]{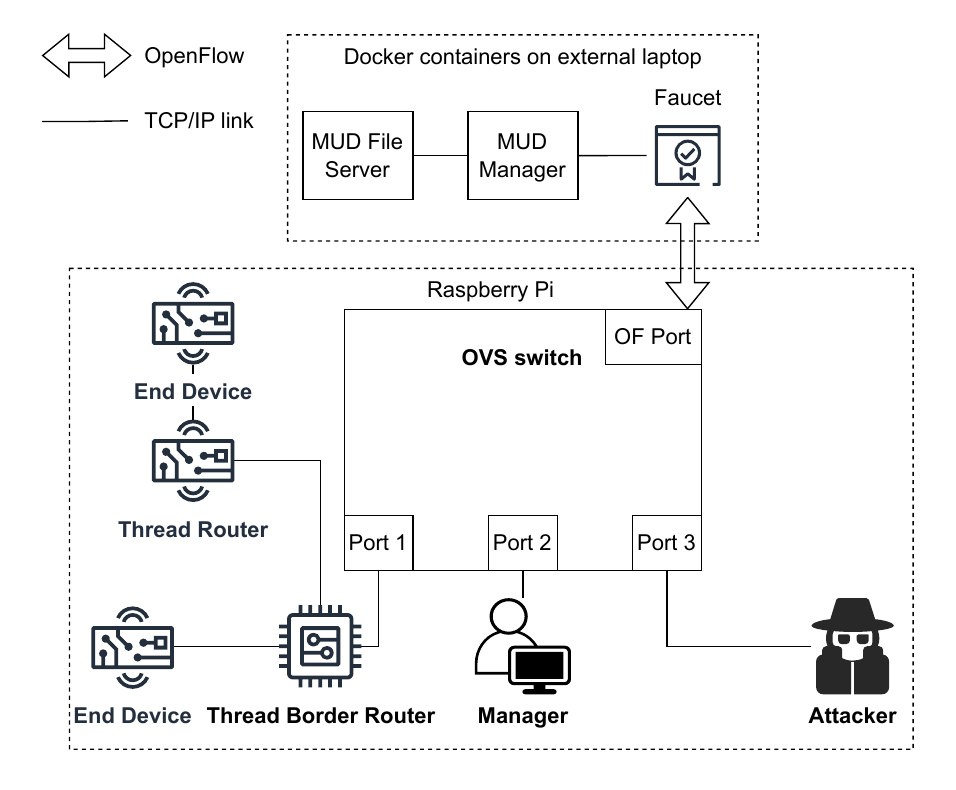}
    \caption{Proof-of-concept architecture, with an administrator and an attacker connected for demonstration purposes.}
    \label{fig:demo-network}
\end{figure}

We implemented and deployed a proof-of-concept of \name, illustrated in Fig.~\ref{fig:demo-network} and described in more detail below.

\medskip
\noindent{\bf Hardware Details.}
We deployed a real-world Thread network using several constrained \gls{IoT} devices provided by Nordic, i.e., nRF52840 and nRF52833 \glspl{SOC}~\cite{nrf52840, nrf52833}. These devices feature a 32-bit ARM® Cortex™-M4 CPU with a floating point unit running at 64 MHz, an IEEE 802.15.4-2006 radio transceiver working at the carrier frequency of 2.4 GHz and a Power Management Unit for low power consumption, allowing them to serve both as \gls{FTD} and \gls{MTD} devices. We also used several Raspberry PIs (models 3 and 4) as Thread border routers~\cite{raspberry-pi3}. 
A Raspberry Pi has considerably more computational resources than both the Thread devices, offering at least a Quad Core 1.2GHz Broadcom BCM2837 64bit CPU and 1 GB of RAM. 
Since Raspberry Pis do not feature an IEEE 802.15.4 capable radio, they require a \gls{RCP} to provide this interface, as provided by the Nordic boards. Finally, we deployed the \gls{SDN} controller, \gls{MUD} Manager, and the external \gls{MUD} file Server as hardware-agnostic Docker containers on a general-purpose laptop.

\medskip
\noindent{\bf Software Details.} Our proof-of-concept used, extended, and integrated together several existing open-source projects.
We used the open-source implementation of the OpenThread protocol stack available at~\cite{openthread-nrf528}, and integrated further such implementation into the Zephyr \gls{RTOS}, which provides a flexible firmware for our Thread devices~\cite{zephyr}. The Zephyr \gls{RTOS} is highly configurable to provide different logging and debugging options as well as different access methods, such as \gls{UART} and Telnet. 
Zephyr provides a custom tool to specify \emph{overlays}, i.e., additions to the existing firmware. Overlays allowed us to extend the default firmware of OpenThread by: (i) storing the MUD URL, (ii) extending the messages \emph{MLE Update Request} and \emph{Child ID Request} of \gls{MLE} to work according to \name, (iii) forcing devices to behave as either \gls{MTD} or \gls{FTD}, and (iv) enabling additional services, e.g., Telnet clients and servers. 

We also used and extended the official implementation of the OpenThread border router available at~\cite{openthread-otbr} to allow communication of the Thread Border Router with the SDN controller. In particular, we extended such implementation through the definition of the IP address of the SDN controller and the inclusion at startup time of an automated route to and from the controller for handling MUD traffic.
%

We deployed \gls{SDN} switches using the \gls{OVS} software~\cite{openvswitch}. For the \gls{SDN} controller, we selected Faucet among the available options, as it was the most actively maintained at the time of writing.
We further customized it by adding a callback that checks incoming OpenFlow \textit{packet\_in} messages to check for \gls{MUD} messages. 
We initially configured the controller to distribute to the SDN switches two base \glspl{ACL}: (i) \textit{allow-mud}, which filters out all \gls{MUD} messages, and (ii) \textit{block-thread}, which blocks all traffic to and from the Thread network.

Our implementation of the \emph{\gls{MUD} Manager} exposes a basic HTTP API with only one endpoint, namely \verb|/enroll|. To notify the MUD Manager that a new device has joined the network, the \gls{SDN} controller sends a \verb|POST| request to the endpoint with a \gls{JSON} payload containing the \gls{OMR} \gls{IP} address of the new device and its \gls{MUD} \gls{URL}. Then, the \gls{MUD} Manager downloads this file and processes it further. Faucet can automatically deploy  \glspl{ACL} in \gls{YAML} format (contained in the MUD file) into forwarding rules. Finally, for some of our experiments, we also realized attacks coming from the external network. To this aim, we implemented some attack scripts for the purpose (see Sec.~\ref{sec:results}) and launched them as standalone scripts, injecting traffic from our laptop (external world) to the Thread network.
We release the source code of our proof of concept open-source at~\url{https://doi.org/10.5281/zenodo.19206286}.

\section{Performance Evaluation} \label{sec:results}
In this section, we present results based on real-world experiments on our proof of concept, demonstrating the effectiveness of \name, in terms of security (correct filtering of unauthorized traffic) and overhead (added network latency and memory requirements).
The code, scripts, and instructions to reproduce our experiments are also available at~\url{https://doi.org/10.5281/zenodo.19206286}.

\subsection{Experiment 1: Malicious incoming traffic \label{sec:results-incoming}}
Many attacks on \gls{IoT} devices exploit their unprotected exposure on the public Internet and the presence of services that remain unnecessarily open. Thus, to evaluate the robustness of our solution and its merit compared with the current state-of-the-art, in this experiment, we investigate the following research question: \emph{To what extent can \name\ prevent incoming malicious traffic from compromising Thread devices, compared to existing MUD-based security solutions for Thread networks?}


\begin{figure}[ht]
    \centering
    \includegraphics[width=.7\columnwidth]{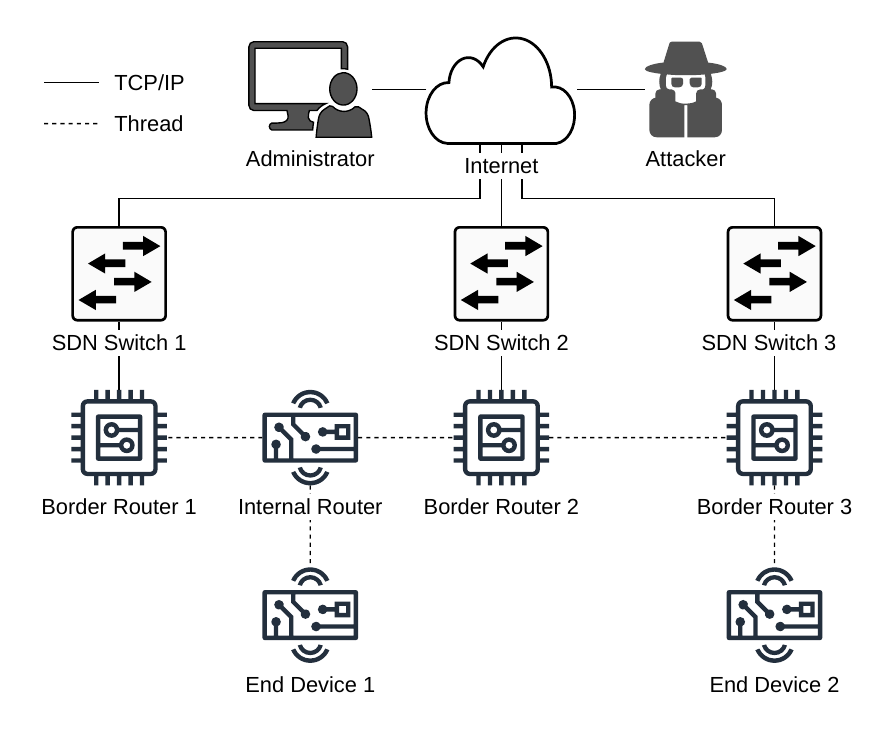}
    \caption{Network layout for Experiment 1.}
    \label{fig:experiment1}
\end{figure}
\medskip
\noindent{\bf Settings.} \Cref{fig:experiment1} presents the network layout considered for this experiment. 
We deployed three Thread border routers, each attached to three \gls{SDN} switches. We configured the border routers to be all part of the same Thread network, including a single internal Thread router and two end devices. Note that one Thread end device is connected to an internal Thread router, but not directly to any border router. Moreover, the end devices can exchange external traffic through different Thread border routers.
All three \gls{SDN} switches are controlled by a single \gls{SDN} Controller and a \gls{MUD} Manager. 
We configured the Thread end devices with a \gls{MUD} \gls{URL} that points to a \gls{MUD} file that considers as \emph{allowed} all \gls{ICMP} packets and all telnet traffic (\gls{TCP} traffic to the device with destination port 23 and from the device with source port 23) with a source or destination \gls{IP} address associated with the domain name \textit{manager.local}. We report in \Cref{lst:short-telnet} the excerpt of the deployed MUD file.

\begin{lstlisting}[caption={A MUD ACL which allows telnet access to the device from the domain name "manager.local"},label={lst:short-telnet}, language=json,firstnumber=1,basicstyle=\ttfamily\footnotesize]
{
    "name": "allow-telnet-to-device-from-manager",
    "type": "ipv6-acl-type",
    "aces": {
        "ace": [
            {
                "name": "allow-telnet",
                "matches": {
                    "ipv6": {
                        "ietf-mud:src-dnsname": "manager.local",
                        "protocol": 6
                    },
                    "tcp": {
                        "ietf-mud:direction-initiated": "to-device",
                        "destination-port": {
                            "operator": "eq",
                            "port": 23
                }}},
                "actions": {
                    "forwarding": "accept"
}}]}}
\end{lstlisting}
To assess how effective \name\ is in enforcing this policy, we implemented and deployed a script mimicking the behavior of the early Mirai malware~\cite{antonakakis2017understanding}. In its first instance, Mirai infected commercial \gls{IoT} devices by exploiting open Telnet ports and weak credentials initially deployed by manufacturers of IoT devices. This led to the creation of a botnet containing over 600.000 devices, with some of them still infected at the time of this writing. In line with such attack logic, we let the attacker and system administrator initiate 100 telnet sessions with each Thread end device on the network. A connection attempt is deemed successful if the \gls{TCP} SYN packet to port 23 is received by the Thread device and the corresponding \gls{TCP} SYN ACK packet is received by the initiating entity. The rationale behind this experiment is that our solution should prevent the compromise of Thread devices, by allowing telnet access only by local \textit{managers} of the devices and not from the external Internet.
\textcolor{black}{On such a network, we deployed and evaluated the effectiveness of both \name\ and the only competing approach in the literature focusing on the application of MUD to Thread networks, i.e., the proposal by Houben et al. in~\cite{thesis_luke_paper}. Note that other solutions from the literature, as discussed in Sec.~\ref{sec:related-work}, cannot be applied here since they rely on protocols unavailable in Thread networks.}

\medskip
\noindent{\bf Results.} We report the results of experiment 1 with \name\ in \Cref{tab:exp1} and with the solution by Houben et al.~\cite{thesis_luke_paper} in \Cref{tab:exp1-mudthread-allow} and~\Cref{tab:exp1-mudthread-block}, considering the option \emph{block by default} disabled and enabled, respectively. We report in the column \emph{Source} the entity originating traffic. The columns \emph{BR} and \emph{End Device} indicate the traffic path, i.e., the IDs of the Thread border router and end device which managed the traffic (see Fig.~\ref{fig:experiment1}). For the experiments with the benchmark in~\cite{thesis_luke_paper}, in line with their implementation and testing, we consider \gls{ICMP} echo requests and replies as allowed traffic (expected to be successful) and Telnet sessions as disallowed one (expected to be unsuccessful).
\begin{table}[tp]
    \centering
    \resizebox{\columnwidth}{!}{
    \begin{tabular}{lcc|cc|cc}
        \multirow{2}{*}{Source} & \multirow{2}{*}{Border Router} & \multirow{2}{*}{End Device} & \multicolumn{2}{c|}{SYN} & \multicolumn{2}{c}{SYN ACK} \\
        & &  & Sent & Received & Sent  & Received \\
        \hline
        Admin & 1 & 1 & 100 & 100 & 100 & 100  \\
        Admin & 1 & 2 & 100 & 100 & 100 & 100  \\
        Admin & 2 & 1 & 100 & 100 & 100 & 100  \\
        Admin & 2 & 2 & 100 & 100 & 100 & 100  \\
        Admin & 3 & 1 & 100 & 100 & 100 & 100  \\
        Admin & 3 & 2 & 100 & 100 & 100 & 100  \\ 
        \hline
        Attacker & 1 & 1 & 100 & 0 & 0 & 0 \\
        Attacker & 1 & 2 & 100 & 0 & 0 & 0 \\
        Attacker & 2 & 1 & 100 & 0 & 0 & 0 \\
        Attacker & 2 & 2 & 100 & 0 & 0 & 0 \\
        Attacker & 3 & 1 & 100 & 0 & 0 & 0 \\
        Attacker & 3 & 2 & 100 & 0 & 0 & 0
    \end{tabular}
    }
    \caption{Results of experiment 1 with \name. Allowed traffic always reaches the end device, while malicious traffic is always blocked.
    }
    \label{tab:exp1}
\end{table}
\begin{table}[tp]
    \centering
     \resizebox{\columnwidth}{!}{
    \begin{tabular}{ccc|cc|cc}
        \multirow{2}{*}{Type} & \multirow{2}{*}{Border Router} & \multirow{2}{*}{End Device} & \multicolumn{2}{c|}{Request} & \multicolumn{2}{c}{Reply} \\
        & &  & Sent & Received & Sent  & Received \\
        \hline
        \gls{ICMP} & 1 & 1 & 100 & 100 & 100 & 100  \\
        \gls{ICMP} & 1 & 2 & 100 & 100 & 100 & 100  \\
        \gls{ICMP} & 2 & 1 & 100 & 100 & 100 & 100  \\
        \gls{ICMP} & 2 & 2 & 100 & 100 & 100 & 100  \\
        \gls{ICMP} & 3 & 1 & 100 & 100 & 100 & 100  \\
        \gls{ICMP} & 3 & 2 & 100 & 100 & 100 & 100  \\
        \hline
        Telnet & 1 & 1 & 100 & 100 & 100 & 100  \\
        Telnet & 1 & 2 & 100 & 100 & 100 & 100  \\
        Telnet & 2 & 1 & 100 & 100 & 100 & 100  \\
        Telnet & 2 & 2 & 100 & 100 & 100 & 100  \\
        Telnet & 3 & 1 & 100 & 100 & 100 & 100  \\
        Telnet & 3 & 2 & 100 & 0 & 0 & 0  \\

    \end{tabular}
    }
    \caption{Results of experiment 1 with the solution by Houben et al.~\cite{thesis_luke_paper}, with the option \emph{block by default} disabled. 
    }
    \label{tab:exp1-mudthread-allow}
\end{table} 
\begin{table}[tp]
    \centering
    \resizebox{\columnwidth}{!}{
    \begin{tabular}{ccc|cc|cc}
        \multirow{2}{*}{Type} & \multirow{2}{*}{Border Router} & \multirow{2}{*}{End Device} & \multicolumn{2}{c|}{Request} & \multicolumn{2}{c}{Reply} \\
        & &  & Sent & Received & Sent  & Received \\
        \hline
        \gls{ICMP} & 1 & 1 & 100 & 0 & 0 & 0  \\
        \gls{ICMP} & 1 & 2 & 100 & 0 & 0 & 0  \\
        \gls{ICMP} & 2 & 1 & 100 & 0 & 0 & 0  \\
        \gls{ICMP} & 2 & 2 & 100 & 0 & 0 & 0  \\
        \gls{ICMP} & 3 & 1 & 100 & 0 & 0 & 0  \\
        \gls{ICMP} & 3 & 2 & 100 & 100 & 100 & 100  \\
        \hline
        Telnet & 1 & 1 & 100 & 0 & 0 & 0  \\
        Telnet & 1 & 2 & 100 & 0 & 0 & 0  \\
        Telnet & 2 & 1 & 100 & 0 & 0 & 0  \\
        Telnet & 2 & 2 & 100 & 0 & 0 & 0  \\
        Telnet & 3 & 1 & 100 & 0 & 0 & 0  \\
        Telnet & 3 & 2 & 100 & 0 & 0 & 0  \\

    \end{tabular}
    }
    \caption{Results of experiment 1 with the solution by Houben et al.~\cite{thesis_luke_paper}, with the option \emph{block by default} enabled.
    }
    \label{tab:exp1-mudthread-block}
\end{table}

From the results in \Cref{tab:exp1}, we highlight the effectiveness of \name: no SYN packet to port 23 originating from the \textit{Attacker} is received by any Thread end device (0\% attack success rate). Furthermore, all SYN packets to port 23 originating from the \textit{Administrator} are forwarded to the end devices, and all their SYN ACK responses are forwarded in turn to the \textit{Administrator}, guaranteeing correct device access. This behavior is verifiably enforced for all tested routes.

Instead, when considering the solution in~\cite{thesis_luke_paper} with the \emph{block-by-default} option disabled, while all \gls{ICMP} echo requests and replies were forwarded correctly, only the telnet traffic to End Device 2 going through Border Router 3 was blocked correctly (last line of~\Cref{tab:exp1-mudthread-allow}). All other Telnet traffic, although being disallowed by the MUD file of the end devices, reached such devices. This is due to the incomplete MUD deployment strategy of the solution in~\cite{thesis_luke_paper}, which enforces MUD-based rules only to the Thread border router directly connected to the end device. An alternative option is to enable the policy \emph{block-by-default}, so that all traffic not directly matching a policy is dropped. As shown in \Cref{tab:exp1-mudthread-allow}, although this policy successfully blocks all Telnet traffic to the devices, it also blocks almost all ICMP traffic that should be allowed instead. Only \gls{ICMP} traffic to End Device 2 through Border Router 3 is forwarded successfully, as they are directly connected. 

Overall, our results demonstrate that \name\ is more effective and secure than the competing solution in~\cite{thesis_luke_paper}, allowing to enforce MUD-based rules in Thread networks with several border routers.


\subsection{Experiment 2: Malicious outgoing traffic \label{sec:restults-outgoing}}
Local attackers with physical access to the Thread devices can physically tamper and compromise them, potentially enabling the delivery of malicious traffic outside the Thread network (e.g., let devices participate in a \gls{DDoS} attack, similarly to Mirai). Therefore, the complete effectiveness of \name\ should be assessed by also checking that our solution can stop such malicious traffic from flooding the public Internet network. This is investigated through the following research question: \emph{To what extent can \name\ prevent outgoing malicious traffic from physically-infected Thread devices?}

\begin{figure}[h]
    \centering
    \includegraphics[width=.7\columnwidth]{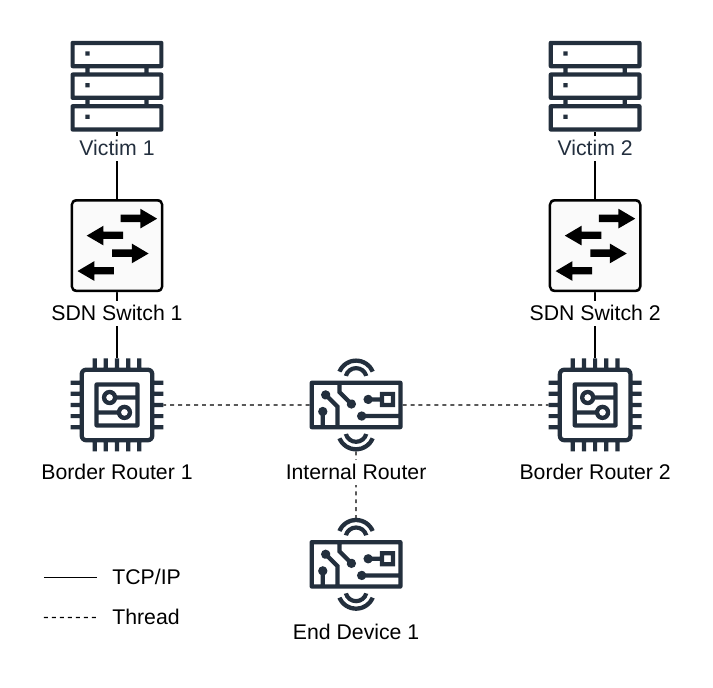}
    \caption{Network layout for experiment 2.}
    \label{fig:experiment2}
\end{figure}

\medskip
\noindent{\bf Setting.} \Cref{fig:experiment2} illustrates the network layout considered for this experiment. 
We deployed two Thread border routers attached to two \gls{SDN} switches. The border routers are all part of the same Thread network, which also includes a single internal router and a single end device. The \gls{SDN} switches are controlled by a single \gls{SDN} Controller and a \gls{MUD} Manager. We configured the Thread end device 1 with a \gls{MUD} \gls{URL} that points to a \gls{MUD} file allowing \gls{ICMP} traffic to be exchanged only with entities corresponding to the domain names \textit{manager.local} and \textit{manager1.local}. This traffic allows us to verify that the network is configured correctly. \gls{ICMP} traffic to any other destination is not allowed. 

To evaluate the effectiveness of the MUD Manager in enforcing this policy, we implemented a script on the Thread end devices that emulates a \gls{DDoS} attack in the form of an \gls{ICMP} flood targeting two external hosts (\textit{Victim~1} and \textit{Victim~2}). The script on End Device~1 generates \gls{ICMP} echo requests toward both victims at increasing rates, 1, 10, and 100 packets per second, for 60 seconds.


\medskip
\noindent{\bf Results.} \Cref{tab:exp2} reports the results of experiment 2.
\begin{table}[tp]
    \centering
    \resizebox{\columnwidth}{!}{
    \begin{tabular}{cc|ccc}
         Victim & Rate (packet/s) & Sent & Received \gls{SDN} switch & Received by Victim  \\
         \hline
         1 &   1 &   60 &   60 & 0 \\
         1 &  10 &  600 &  600 & 0 \\
         1 & 100 & 6000 & 2931 & 0 \\
         2 &   1 &   60 &   60 & 0 \\
         2 &  10 &  600 &  600 & 0 \\
         2 & 100 & 6000 & 4480 & 0
    \end{tabular}
    }
    \caption{Results of experiment 2. All malicious traffic that reached the SDN switches is successfully blocked.}
    \label{tab:exp2}
\end{table}
First, we observe that no packets were received by the victim nodes in any of the performed attempts, indicating that all malicious traffic that reached the SDN switches was successfully blocked. These results show that even if a device is physically compromised after joining the network, \name\ effectively mitigates follow-up attacks to external networks (e.g., the Internet). We also observe that when the rate at which the end device delivers packets increases significantly (100 packets/sec), the SDN switch receives only a fraction of those packets. This is not an issue due to \name, but rather a physical limitation of Thread, which uses a low-power and lossy communication technology (IEEE 802.15.4) supporting a limited traffic rate. Thus, \name\ is also robust to the traffic rates sustainable by commercial Thread networks.

\subsection{Experiment 3: Latency Overhead \label{sec:restults-impact}}
\name\ requires each Thread border router to connect to an SDN switch, which may increase the experienced network latency. Thus, we performed an experiment aiming to answer the following research question: \emph{What is the impact of \name\ on the performance of the Thread network?}

\begin{figure}[tp]
    \centering
    \subfloat[Network scenario 1: the Thread network consists of one border router and one end device.]{
        \includegraphics[width=.65\columnwidth]{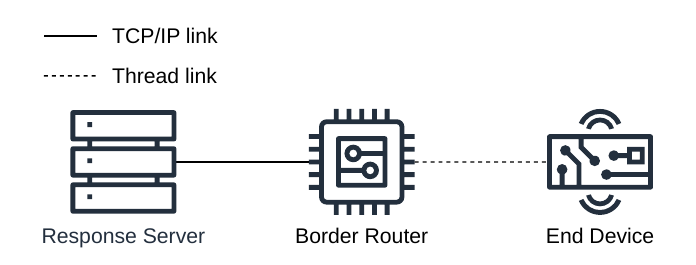}
        \label{subfig:exp3-1}
    }\\[1ex]
    \subfloat[Network scenario 2: the network includes one SDN switch, but no MUD Manager and no ACL enforcement.]{
        \includegraphics[width=.65\columnwidth]{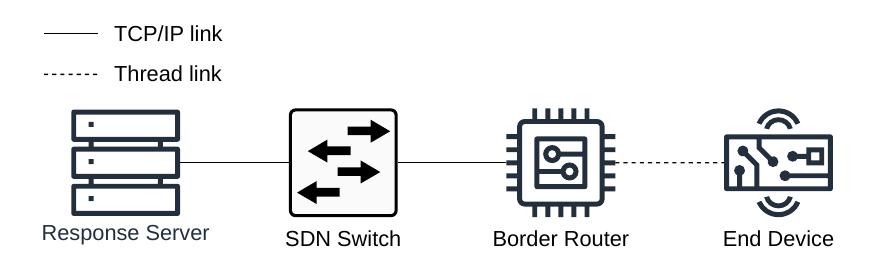}
        \label{subfig:exp3-2}
    }\\[1ex]
    \subfloat[Network scenario 3 and 4: the network includes one SDN switch with MUD-based ACL enforcement.]{
        \includegraphics[width=.65\columnwidth]{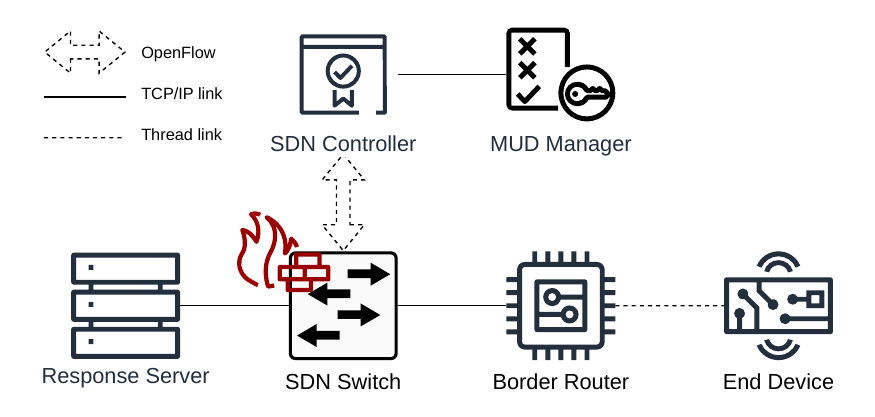}
        \label{subfig:exp3-3}
    }
    \caption{Network layouts for experiment 3.}
    \label{fig:experiment3}
\end{figure}

\medskip
\noindent{\bf Setting.} \Cref{fig:experiment3} illustrates the three network scenarios considered for this experiment.
Scenario 1 (\Cref{subfig:exp3-1}) consists of a single Thread end device, one Thread border router, and a remote server located one hop away from the border router. Scenario 2 (\Cref{subfig:exp3-2}) features one \gls{SDN} switch (not enforcing MUD rules) located between the border router and the remote server, allowing us to evaluate the impact of the \gls{SDN} switch on network latency. Scenario 3 (\Cref{subfig:exp3-3}) adds to Scenario 2 also MUD-based rules enforcement on the SDN switch, allowing us to evaluate the impact of the enforcement of such rules. Finally, Scenario 4 considers the same scenario as in \Cref{subfig:exp3-3}, but with an increased number of rules. Specifically, we increase the size of the \gls{ACL} with several non-matching rules. While it might be possible to group firewall rules between Thread devices, this falls outside of the scope of this paper. Therefore, we consider the worst-case scenario where each new Thread device adds several rules that only permit traffic to or from that device. 
For comparison, we also consider a baseline measurement of the \gls{RTT} between two directly connected Thread nodes. 

\begin{figure}[tp]
    \centering
    \includegraphics[width=0.75\columnwidth]{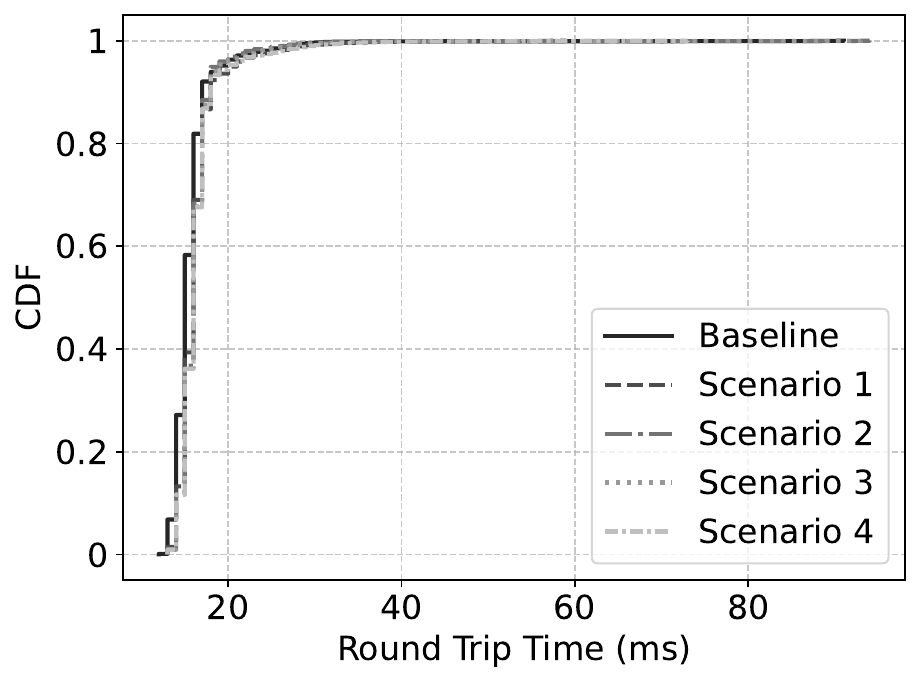}
    \caption{\textcolor{black}{Results of experiment 3. All scenarios report similar Cumulative Distribution Function (CDF), except the baseline (no Thread router), confirming that \name\ introduces negligible latency overhead.}
    }
    \label{fig:ex3-results}
\end{figure}
\medskip
\noindent{\bf Results.} \Cref{fig:ex3-results} presents that the results of all four considered scenarios are very similar, with median \gls{RTT} values of $\approx$ 16 ms. Only the baseline scenario (no Thread border routers) shows values which are, on average, $1$~ms lower than the other investigated scenarios, including Scenario 1, which does not include SDN switches at all. \textcolor{black}{This confirms that \name\ introduces negligible latency on network communications, emerging as a lightweight solution even in a constrained network.}


\subsection{Experiment 4: Scalability}
\label{sec:exp_scalability}

As the number of border routers in the Thread network increases, the number of flow rules managed by the SDN controller and switches increases, potentially affecting the overhead on the network. To investigate how scalable \name\ is with the increasing size of the Thread network, we formulate the following research question: \emph{What is the impact of an increasing number of Thread Border Routers on the number of flows and ACL rules managed by the SDN switches?}


\medskip
\noindent{\bf Setting.} The network scenario considered for this experiment aligns with scenario 3 of experiment 3 (see \Cref{subfig:exp3-3}), with one Thread end device, one SDN switch per Thread Border Router, one SDN controller, and a MUD Manager. In our proof-of-concept implementation, the MUD file of the Thread end device generates 8 flow rules. We then deployed an increasing number of Thread border routers, up to 10. Since the configuration of the flow rules given by the SDN controller to the SDN switches depends on the number of switches, increasing the number of Thread border routers (i.e., SDN switches) increases the number of flow rules stored at such switches. To scale up the number of SDN switches without deploying additional hardware, we ran additional instances of \gls{OVS} and delivered from each of them a \gls{UDP} message containing the \gls{MUD} \gls{URL} of a fictitious Thread device.

\begin{figure}[tp]
    \centering
    \subfloat[Number of flow rules before and after the MUD file of a new device is added to the SDN switch, with a given number of Thread border routers.]{
        \includegraphics[width=.7\columnwidth]{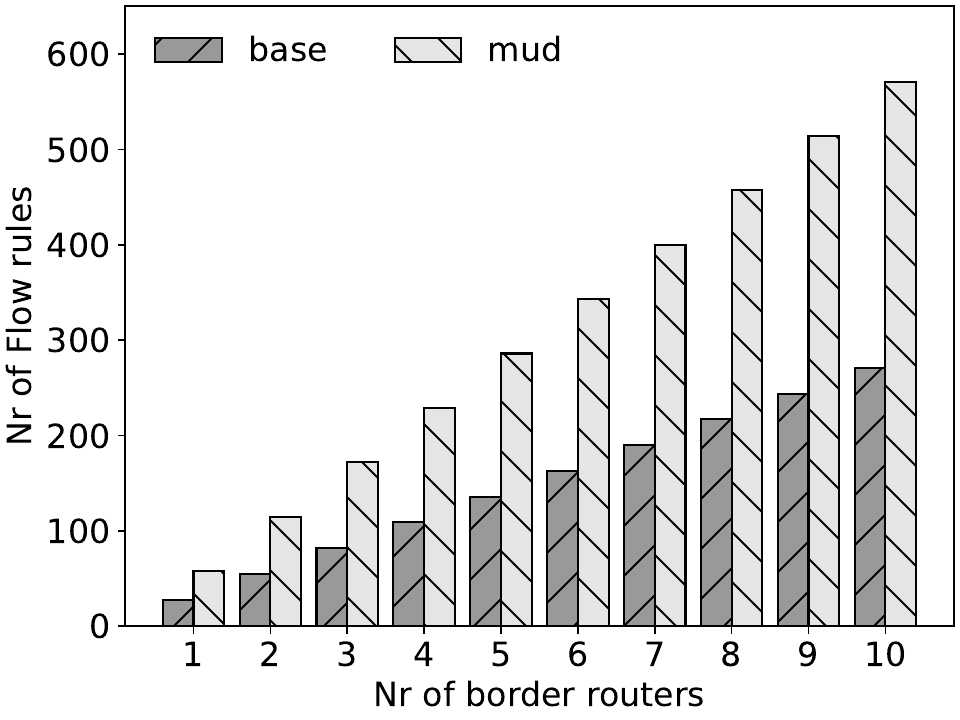}
        \label{fig:exp4_flow_rules}
    }
    \\[1ex]
    \subfloat[Number of ACL rules required by \name, with a given number of Thread border routers.]{
        \includegraphics[width=.7\columnwidth]{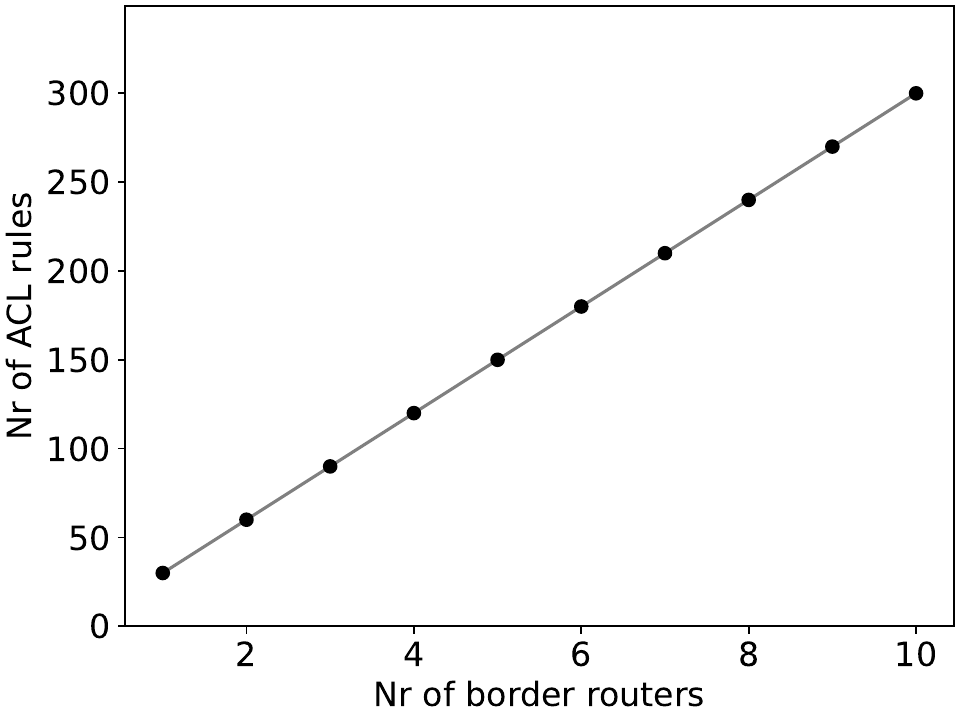}
        \label{fig:exp4_acl_rules}
    }
    \caption{Results for experiment 4: The number of flow rules per border router added by the MUD file of a new device.}
    \label{fig:exp4_results}
\end{figure}

\medskip
\noindent{\bf Results.} \Cref{fig:exp4_flow_rules} reports the number of flow rules per \gls{SDN} switch by increasing the number of deployed \gls{SDN} switches, while \Cref{fig:exp4_acl_rules} further shows the number of deployed flow rules. 
Overall, our results report a linear increasing trend of the rules installed on the SDN switches with the number of deployed border routers. We can explain this trend by looking at how Faucet translates the \glspl{ACL} of the \gls{MUD} file into flow rules.
Consider \glspl{ACE} extracted from the MUD file shown in \Cref{lst:short-telnet}. When only one \gls{SDN} switch and one Thread border router are configured, the MUD file results in the three flow rules shown in \Cref{lst:flow-rules}.
\begin{lstlisting}[float,caption={Flow rules which allow telnet access from the IP range fd99:aaaa:bbbb:200::/64},label={lst:flow-rules}]
table=1, tcp6, dl_vlan=100, ipv6_src=fd99:aaaa:bbbb:200::/64,
    ipv6_dst=fd71:666b:b2e1:bfd9:44ea:9c0a:d388:64ae, tp_dst=23 actions=goto_table:2
table=1, tcp6, dl_vlan=200, ipv6_src=fd99:aaaa:bbbb:200::/64,
    ipv6_dst=fd71:666b:b2e1:bfd9:44ea:9c0a:d388:64ae, tp_dst=23 actions=goto_table:2
table=1, tcp6, dl_vlan=300, ipv6_src=fd99:aaaa:bbbb:200::/64,
    ipv6_dst=fd71:666b:b2e1:bfd9:44ea:9c0a:d388:64ae, tp_dst=23 actions=goto_table:2    
\end{lstlisting}
The rules differ only in the field \verb|dl_vlan|, which distinguish the three ports configured for the switch. When considering additional switches, these rules are duplicated for each \gls{SDN} switch. Thus, considering \gls{SDN} switches with the same number of ports $p$ and denoting as $A$ the set of all \glspl{ACE} in the \gls{MUD} file and $B$ the set of all deployed border routers, we can compute the number of rules $n$ added by a single \gls{MUD} file as $n = (A + 2) \cdot p \cdot B$, where the two additional rules added to the number of \glspl{ACE} are two "deny by default" rules that block all messages that do not match any other \gls{ACE} in the \gls{MUD} file. We can obtain the results in \Cref{fig:exp4_acl_rules} by using $p=3$ and $A=8$, resulting in the linear equation $n=30 \cdot B$.

\textcolor{black}{In summary, our analysis shows analytically and experimentally the linear memory and processing overhead of \name. Thanks to the evaluation, network administrators can plan ahead: they can precompute the overhead resulting from the integration of \name\ based on the network size and properties, and tailor the deployments of applications running on top of it based on the expected additional overhead.}


\section{Discussion} \label{sec:discussion}

\noindent{\bf Implications on research and practice.} The results provided in Sec.~\ref{sec:results} demonstrate that \name\ advances the current state of the art by providing a solution allowing to fully secure constrained Thread networks against attacks exploiting unused protocols, without any assumption on the size and distribution of the networks. While existing approaches are effective only for small networks, \name\ is robust also with several Thread border routers, in large mesh networks typical of deployments in buildings and large geographic areas (e.g., woods, agricultural fields), as demonstrated by the results of experiments 1 and 2. \name\ also emerges as a lightweight solution (experiment 3), whose memory overhead scales linearly with the size of the network (experiment 4). These overhead factors should be carefully taken into account by practitioners and system administrators deploying the system and developing applications on top of it in the real world. 
\textcolor{black}{Overall, \name\ demonstrates the feasibility of extending MLE in a standard-compliant way to fully support MUD in Thread networks. Similar to standardized extensions for TCP/IP-based IoT networks, it is possible to standardize MLE and use our solution as an additional MUD URL delivery method, tailored for low-power IoT networks. Prior to standardization, network administrators should update Thread devices to support our manufacturer-independent extension.
}

\medskip
\noindent{\bf Limitations.} We considered attackers who compromise only the Thread devices (end devices, routers). More advanced attackers could take over the SDN switches or controllers, being able to affect the correct functionality of our solution. \textcolor{black}{A single compromised SDN switch or Thread border router could advertise poisoned routes to Thread end devices, and compromise the security of \name. We do not address this problem, but lightweight solutions are available to solve this problem, as discussed in~\cite{zhou2018_ton} and~\cite{dargahi2017_comst}, to name a few. Such solutions can be integrated out of the box to further strengthen the overall network security.}

Moreover, we do not consider attacks exploiting vulnerabilities of protocols used by the Thread devices as part of their operation, in line with the rationale of the \gls{MUD} specification. Attackers exploiting such vulnerabilities would successfully bypass our solution, as well as any solution relying solely on MUD. Therefore, to increase the security guarantees against such adversaries, additional security systems are necessary to monitor the exploitation of such vulnerabilities. 

\textcolor{black}{Finally, we note that our work focuses on the integration of MUD with the Thread protocol suite. Other IoT technologies likely use a different protocol stack; thus, to run on such networks, \name\ should be redesigned. In this regard, \name\ and its design methodology provide a blueprint for integrating MUD-based network access control in any IoT network not running the TCP/IP protocol stack, thus potentially creating impact beyond solely Thread.}

\section{Conclusion}\label{sec:conclusion}
This paper proposed \name, a lightweight and scalable framework for enforcing MUD-based network access control in a constrained large-scale Thread-powered IoT network. 
\name\ extends the \gls{MLE} protocol to allow for the scalable delivery of MUD-related information by any Thread end nodes, even when such nodes are not directly connected to a Thread border router. Moreover, \name\ accounts for the deployment of multiple border routers, and synchronizes network access control policies through the integration of the \gls{SDN} technology. 
We experimentally tested the security and overhead of \name\ through a proof-of-concept deployment using actual Thread devices running the OpenThread protocol stack and constrained border routers running on Raspberry Pi devices. Our tests experimentally demonstrate the effectiveness of \name, as its deployment blocks all tested attacks while allowing regular IoT traffic, in contrast to existing solutions. Moreover, we demonstrate the limited overhead of our solution, scaling linearly with the number of deployed border routers. In future work, we plan to integrate \name\ within an intrusion detection system for Thread to further increase network security against attacks that exploit vulnerabilities in legitimate protocols.



\section*{Acknowledgements}
\label{sec:acks}
This research was made possible by the INTERSECT project, Grant ID NWA.1162.18.301, funded by the Netherlands Organisation for Scientific Research (NWO) and by the European Network of Excellence dAIEDGE under Grant Agreement Nr. 101120726. The contents herein are solely the responsibility of the author(s). Dominik Roy George contributed while he was at Eindhoven University of Technology.
\balance
\bibliographystyle{ieeetr}
\bibliography{references}




\end{document}